# Demonstration of magnetic and light-controlled actuation of a photomagnetically actuated deformable mirror for wavefront control

**Amit Kumar Jha,[a,b,*] Ewan S. Douglas,[b] Meng Li,[c,d]
Corey Fucetola,[c,e] and Fiorenzo G. Omenetto[c]**
[a]University of Arizona, Wyant College of Optical Sciences, Tucson, Arizona, United States
[b]University of Arizona, Department of Astronomy and Steward Observatory,
University of Arizona Space Astrophysics Lab, Tucson, Arizona, United States
[c]Tufts University, Silk Lab, Department of Biomedical Engineering, Medford,
Massachusetts, United States
[d]Physical Intelligence, Max Planck Institute for Intelligent Systems, Stuttgart, Germany
[e]Sublamit Laboratories, Somerville, Massachusetts, United States

**Abstract.** Deformable mirrors (DMs) have wide applications ranging from astronomical imaging to laser communications and vision science. However, they often require bulky multi-channel cables for delivering high power to their drive actuators. A low-powered DM, which is driven in a contactless fashion, could provide a possible alternative to this problem. We present a photomagnetically actuated deformable mirror (PMADM) concept, which is actuated in a contactless fashion by a permanent magnet and low-power laser heating source. We present the laboratory demonstration of prototype optical surface quality, magnetic control of focus, and COMSOL simulations of its precise photocontrol. The PMADM prototype is made of a magnetic composite (polydimethylsiloxane + ferromagnetic $CrO_2$) and an optical-quality substrate layer and is 30.48 mm × 30.48 mm × 175 $\mu$m in dimension with an optical pupil diameter of 8 mm. It deforms to 5.76 $\mu$m when subjected to a 0.12-T magnetic flux density and relaxes to 3.76 $\mu$m when illuminated by a 50-mW laser. A maximum stroke of 8.78 $\mu$m before failure is also estimated considering a 3× safety factor. Our work also includes simulation of astigmatism generation with the PMADM, a first step in demonstrating control of higher order modes. A fully developed PMADM may have potential application for wavefront corrections in vacuum and space environments. © *The Authors. Published by SPIE under a Creative Commons Attribution 4.0 International License. Distribution or reproduction of this work in whole or in part requires full attribution of the original publication, including its DOI.* [DOI: 10.1117/1.OE.60.12.124102]

## 1 Introduction

Dynamical optical elements that modify wavefronts (phase or amplitude) have a wealth of applications. Examples include nanometer or picometer wavefront correction with deformable mirrors (DMs) for exoplanet imaging,[1,2] digital micromirror devices for projection displays,[3] spatial light modulators for lithography,[4] intracavity use for high-power thin disk resonators,[5] speckle reduction in laser picoprojectors,[6] retinal imaging and vision correction,[7] and image slicing,[8,9] to name a few. Of all these, DMs are of particular interest as they have high actuator count, high reflectivity, and precise actuator movement for both low- and high-order wavefront corrections. However, they often come with significant actuator and drive electronics challenges. For example, common DMs such as microelectromechanical systems (MEMS)[10,11] and lead-magnesium-niobate DMs require hundreds to thousands of channels at >100 V,[12–15] leading to complexity in cable management, voltage supply, and actuation process.

---

*Address all correspondence to Amit Kumar Jha, amitjha074@email.arizona.edu

Despite recent progress[16] at the ∼1000 actuator mark, the requirement of high-voltage controller electronics complicates the scalability of these devices, particularly for vacuum and space applications. Numerous alternatives have recently been developed. For example, to control low-order aberrations, liquid deformable lenses have been tested for space communication applications.[17] However, it still requires high voltage (∼100 V) to control deformation. Magnetic fluid deformable mirrors (MFDMs) provide high surface quality, high precision, and are actuated using magnetic fields generated by low current carrying solenoids but they are limited in degree of freedom,[18–20] as they are constrained to operate horizontally under gravity. Mirrors coated with magnetostrictive material,[21,22] magnetic polymers,[23] and magnetic composite materials[24] eliminate the constraint problem but require more than one magnet or coiled actuators for precise actuation control. The magnetic membrane-based ALPAO mirrors[25] comes with significant development in the actuator count (∼3228) and is driven by low voltage but require high power per actuator (>1 W).[26] Several ground-based observatories[27–30] have implemented highly efficient deformable secondary mirrors with thin shells driven by voice coil actuators for adaptive optics and high-contrast imaging applications.[30] However, due to the high power needed by these actuators, developing a low-powered prototype with high actuator count (∼1000) is challenging while maintaining other critical performance values, such as surface quality and low rate of actuator failure.[31] Large DMs with low-powered actuator systems are currently being studied and actively developed to be tested as on-sky adaptive secondary mirrors[32] to mitigate these issues.

DMs are highly effective in correcting higher order aberrations, however, when it comes to focusing control, fewer actuators are required and several active optic technologies such as autofocus lenses,[33] liquid crystal modulators,[34] and spatial light modulators[35] are commonly used in place of high-powered DMs. For example, the Very Large Telescope interferometer mode uses an air-pressure-driven stainless steel focusing optic instead of a DM architecture for achieving high dynamic range.[36] Thus a gap exists for a deformable optics technology, which not only has a desirable actuator count but is scalable and requires low power/voltage for focusing applications in microscopy, laser communication, retinal imaging, and vacuum- and space-based experiments, for example, in space high-contrast imaging[1] where both power and volume needed for an active optics instrument is limited and of key importance.

DMs that are optically rather than electromechanically driven present a suitable alternative to these issues. Photocontrolled deformable mirrors (PCDMs), where actuation is varied by the intensity of incident light rather than current or voltage, allow optical addressing with the potential for high-density remote control. Several authors have proposed PCDMs[37–40] employing a photo-conductive substrate to vary the voltage applied to a membrane mirror.

Photomagnetic actuators that replace the bias voltage of photoconductive architectures with a magnetic field offer another potential alternative. Recently, Li et al.[41] developed a magnetic composite material with a low Curie temperature tuned to readily demagnetize when subjected to a small heat flux, for example, when illuminated by a laser source. This allows controlling the deflection of the material by demagnetizing it with a low-power laser source without varying the magnetic field. They demonstrated a 66.7 wt. % polydimethylsiloxane (PDMS) and 33.3 wt. % $CrO_2$-based magnetic composite with a cantilever beam architecture getting actuated by a magnet and showing almost full relaxation when heated by a laser source. This material has several applications, ranging from soft robotics[42] to optomechanics.[43] We recently developed a COMSOL model[44] that simulates the same material and estimates the amount of deflection and relaxation caused by the magnet and the laser source reported in Li et al.

In this work, we apply the same magnetic composite material to an optical quality substrate to develop and demonstrate focus control as a proof-of-concept for a photomagnetically actuated deformable mirror (PMADM). The actuation of the optical surface is shown through measurements using magnetic loading and then compared to COMSOL simulation results to further validate modeled extension to photoactuation. Active focus control is the first aberration for correction by any deformable optics and our main goal in this work will be to demonstrate that using the PMADM prototype. In Sec. 2, we describe the methods adopted to simulate, fabricate, and test the PMADM prototype. Section 3 shows the COMSOL simulation results and experimental results obtained from laboratory testing, and Sec. 4 summarizes the main findings of this paper and highlights some of the possible future work.

## 2 Methods

In this section, we describe the methods adopted to simulate, fabricate, and experimentally test the PMADM prototype. Initially, a $CrO_2$ + PDMS composite was directly coated with reflective gold, which failed to provide a sufficiently smooth surface for interferometric testing. Thus a device composed of an optical substrate coated with $CrO_2$ + PDMS was developed. A thin plate's rigidity goes as thickness cubed; so a 25-$\mu$m Si wafer was chosen as a readily available optical substrate to minimize actuation force and provide robustness to the prototype while maintaining a smooth surface for interferometric testing.

### 2.1 Modeling

We have used COMSOL Multiphysics[45] software to develop the finite-element method (FEM) model of our PMADM. The structure is composed of two layers: a magnetic PDMS (PDMS + $CrO_2$) layer (150 $\mu$m) and a Si layer (25 $\mu$m), respectively. Due to the bimorph nature[41,44] of the magnetic layer, it is further divided into two equally thick layers for simulation purposes. Thus the model is constructed with three layers: a PDMS layer with no $CrO_2$ (P-layer with thickness of 75 $\mu$m), a PDMS layer with high $CrO_2$ concentration (C-layer with thickness of 75 $\mu$m), and a silicon layer (Si-layer with thickness of 25 $\mu$m).

The material properties used for the magnetic composite layer have been experimentally determined in the previous works,[41,46] whereas the material property for the Silicon layer is taken from COMSOL Multiphysics built-in materials library. All the values are summarized in Table 1.

The magnetic properties of the $CrO_2$ + PDMS composite layer are a function of its temperature.[44,47] Equation (1) shows the dependence of magnetization ($\vec{M}$) of the magnetic layer on Curie temperature ($T_C$) and substrate temperature ($T$), where $C$ and $\beta$ are the Curie constant and the critical exponent factor with values $C = 5.661$ and $\beta = 0.2984$, respectively,

$$|\vec{M}(T)| = C(T_C - T)^\beta. \tag{1}$$

Equation (2) relates the magnetization ($\vec{M}$) with magnetic susceptibility ($\chi$), permeability of free space ($\mu_0$), and magnetic field density ($\vec{B}$):

$$\vec{M} = (\chi/\mu_0)\vec{B}. \tag{2}$$

Li et al. have shown [see Appendix, Fig. 16(a)] that the slope of the hysteretic magnetization of pretreated $CrO_2$ is nonlinear with applied magnetic field strength. The slope is directly proportional to the magnetic susceptibility ($\chi$); hence, we have used relative permeability of the magnetic PDMS layer as a function of magnetic field strength in our COMSOL model. The estimate for the values has been calculated using the experimental values presented by Li et al and is shown in Fig. 16(b).

**Table 1** Material properties of PDMS (crosslinkage PDMS polymer), magnetic PDMS (crosslinking PDMS with 33.3% $CrO_2$), and Si layer of the PMADM.

| Material properties | P-layer (PDMS) | C-layer ($CrO_2$ + PDMS) | Si-layer (Si) | Units |
|---|---|---|---|---|
| Young's modulus | 2.44 | 0.22 | $1.7 \times 10^5$ | MPa |
| Poisson's ratio | 0.49 | 0.45 | 0.28 | — |
| Density | 0.96 | 1.4 | 2.329 | g/cm$^3$ |
| Thermal conductivity | 0.2 | 0.25 | 130 | W/(m · K) |
| Heat capacity at constant pressure | 2174 | 1840 | 700 | J/(kg · K) |
| Coefficient of thermal expansion | $1.88 \times 10^{-4}$ | $1.588 \times 10^{-4}$ | $2.6 \times 10^{-6}$ | 1/K |

When the substrate is subjected to a magnetic field, it experiences a magnetic load that is governed by the electromagnetic force density equation,[48] given by the following equation:

$$\vec{F}(\vec{r}, t) = (\vec{m} \cdot \vec{\nabla})\vec{B}(\vec{r}, t), \quad (3)$$

where $\vec{F}(\vec{r}, t)$ is the force acting on the PMADM structure, $\vec{m}$ is the magnetic moment, and $\vec{B}(\vec{r}, t)$ is the magnetic field density. Equations (1) and (3) show the dependence of magnetic properties of the PMADM on both the temperature and applied magnetic field strength. Therefore, one can vary the magnetic load on the PMADM by changing the applied magnetic field strength or by changing its temperature, allowing control over the magnitude of deformation.

For a simple case, we used Gaussian laser source in COMSOL to heat the magnetic PDMS side of the PMADM using the following equation:

$$I_{\text{source}}(r) = P_{\text{laser}} \left( \frac{1}{2\pi\sigma^2} \exp\left(-\frac{r^2}{2\sigma^2}\right) \right), \quad (4)$$

where $P_{\text{laser}}$ is the laser power and $\sigma = 2$ mm. The incident laser power heats the area within the optical pupil diameter ($D = 8$ mm) of the PMADM causing its temperature to increase. An increase in temperature leads to a decrease in the magnetization of the magnetic PDMS layer [governed by Eq. (1)] of the PMADM, which causes relaxation in the deformation [governed by Eq. (3)]. To add the convective heat loss in our model, we have also used the heat transfer coefficient, light absorption coefficient, and ambient temperature values. The heat transfer coefficient and the light absorption coefficient for our model is 49.21 W/m² K and 0.97, respectively. The ambient temperature is set to 303 K (room temperature) in our FEM model. Figure 1 shows the illustration and the deformation results (COMSOL simulation) demonstrating the effects of the magnetic loading and the laser heating on the PMADM. Figure 2 shows the relationship between the magnetization of the composite material and its temperature, indicating demagnetization with increasing temperature.

### 2.2 Fabrication

The pretreatment process of $CrO_2$ for preparing the magnetic PDMS is described in detail by Li et al. Once the magnetic PDMS mixture with 66.7 wt. % PDMS and 33.3 wt. % $CrO_2$ is prepared, it is coated to a thickness of ~150 μm onto a silicon wafer with dimensions 30.48 mm × 30.48 mm × 25 μm. A glass slide was translated parallel to the wafer surface with spacers on both sides to evenly apply the viscous magnetic PDMS. This process allows the settling of an even layer of magnetic PDMS over the silicon wafer. Once the coating process is finished, the coated wafer is left on a 150°C hot plate for 20 min to cure the PDMS. After curing, the coated silicon wafer is attached to the acrylic frames to provide enough rigidity to the structure for handling and metrology measurements in the laboratory settings. Figure 3 shows a snapshot of the PMADM taken during the testing phase in the laboratory.

### 2.3 Magnetic Deformation Test

Figure 4 shows the experimental setup used to subject magnetic loading on the PMADM. We have used a stack of eight KJ DC2E[49] disk magnets (hereafter referred to as "magnets") with the north pole facing the magnetic PDMS side (P-layer) of the PMADM. On the silicon wafer side (Si-layer), we have set up a Phasecam 6000 4D interferometer[50] with a collimating lens assembly to measure the surface deformation. The collimating assembly was included to allow compensation for initial focus error introduced by the initial deformation before laser heating, if required. By varying the relative distance between the magnets and the PMADM structure, we changed the magnetic flux density and proceeded to measure the deformation at the Si-layer of the PMADM. Although the magnetic loading test presented here could also be implemented using an electromagnet, displacing a fixed magnet provided an easily modeled, highly repeatable magnetic field.

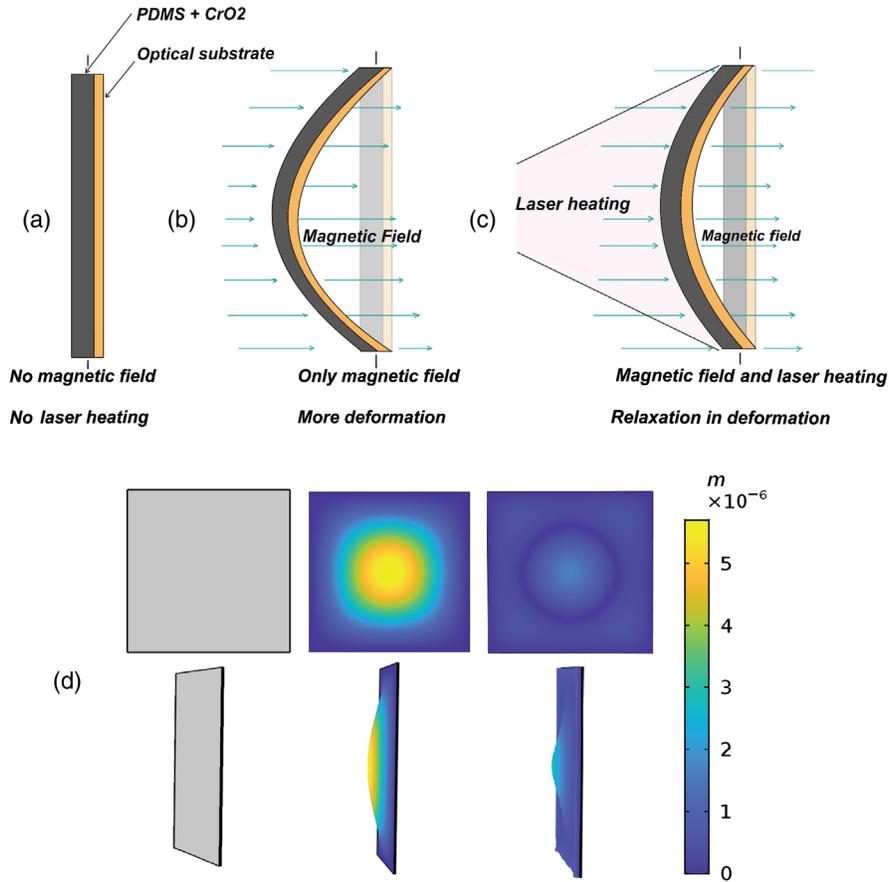

**Fig. 1** (a) Illustration showing the effects of (b) magnetic field and (c) laser heating on the PMADM. The magnetic field causes more deformation when the laser is off. The laser heats the PMADM surface causing demagnetization of the magnetic layer that results in relaxation. COMSOL simulation results of deformation at (d) Si-layer and cross-section view of the PMADM. All the edges of the PMADM are fixed.

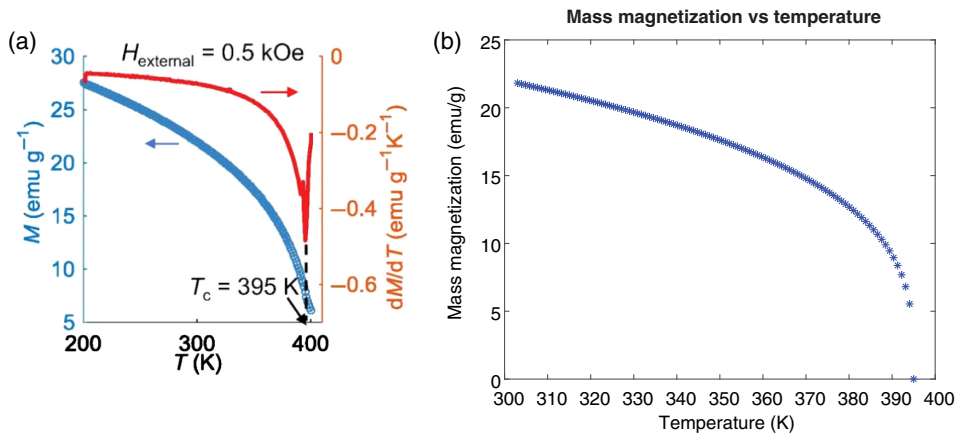

**Fig. 2** (a) Plot of mass magnetization versus temperature taken from Fig. 2D of Li et al.[42] (b) Plot of mass magnetization versus temperature governed by Eq. (1) taken from Fig. 2 of Jha et al.[45] The relationship describes the decreasing magnetization properties of the composite structure with increasing temperature. The magnetization comes to zero when the temperature is equal to the Curie temperature.

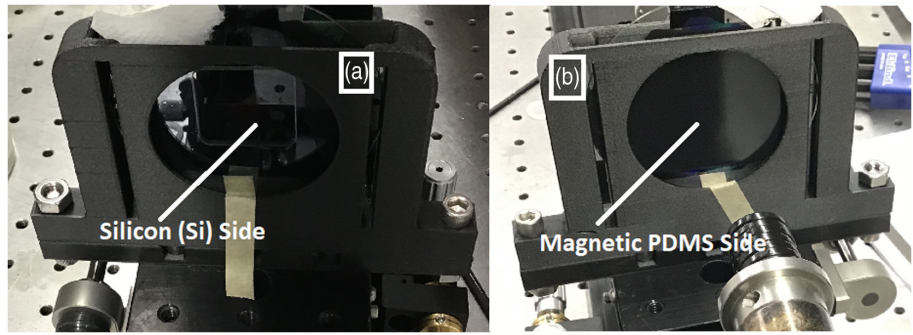

**Fig. 3** Snapshots of PMADM taken in the laboratory. (a) The silicon layer (Si) side and (b) the magnetic PDMS layer side of the structure.

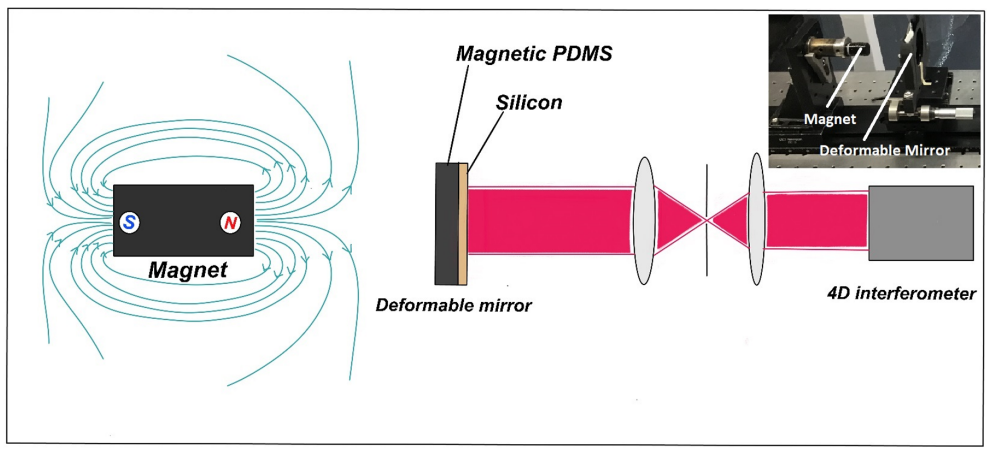

**Fig. 4** Illustration showing the experimental setup used for measuring the surface deformation at the silicon (Si) layer side of PMADM. The image at the top right corner is a snapshot of the laboratory setup showing the magnets and the PMADM on a moving stage. By changing the relative distance between the magnets and the PMADM sample, we can vary the magnetic load subjected on the PMADM structure.

## 3 Results

We present the simulation and experimental results obtained from our COMSOL model and laboratory testing of the PMADM. The COMSOL simulation data have been further post-processed using SAGUARO[51,52] [Software Analysis graphical user interface (GUI) from the University of Arizona for Research in Optics], an open-source MATLAB[53] GUI-based application that uses surface maps obtained from FEM simulation software to calculate optical metrology information. We have used it to calculate the Zernike coefficient values and surface root-mean-square (RMS) figure for our PMADM. The experimental data obtained from the 4D interferometer have been processed in 4SIGHT[50] software to measure metrology information, such as Zernike coefficients.

### 3.1 *Simulation of Magnetic Deformation of PMADM*

In Fig. 5, we have shown the COMSOL simulation of eight magnets[49] delivering the magnetic load on the magnetic side (C-layer) of the PMADM. The separation between the magnets and the PMADM is 12 mm. Figure 5 also shows that the stack of magnets generates a maximum magnetic flux density of around 0.12 T at the magnetic side (C-layer) of the PMADM.

In Fig. 6(a), we have shown the simulated displacement map of the PMADM structure generated due to the magnetic loading in COMSOL. Since the layers are thin, both the P and C layers

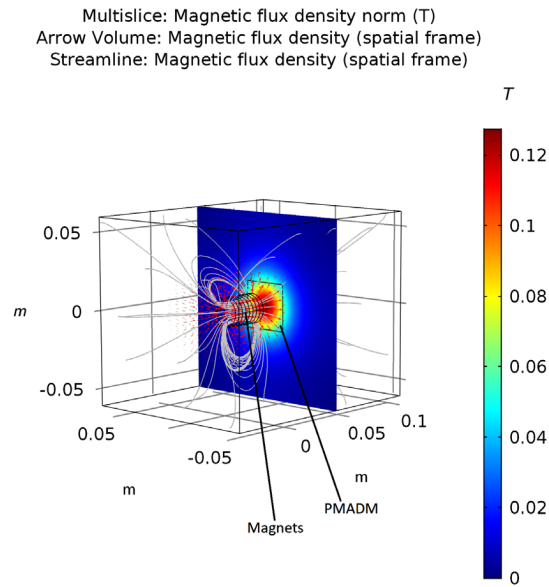

**Fig. 5** COMSOL simulation of eight magnets subjecting a magnetic load on the PMADM structure. The magnetic field lines (in gray) and the magnetic flux density at the C-layer of the PMADM are shown. The separation between the magnet and the PMADM is 12 mm, and a maximum flux density of 0.12 T can be seen at the C-layer of the PMADM.

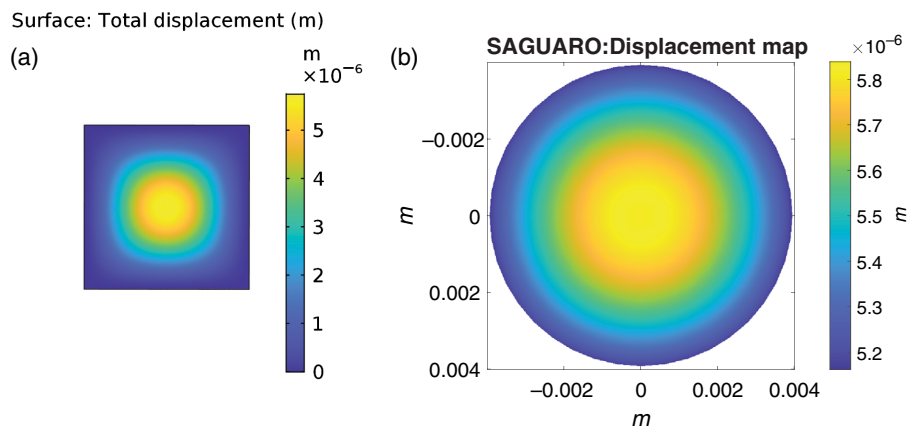

**Fig. 6** (a) Displacement due to the magnetic load on the Si-layer of the PMADM. (b) SAGUARO displacement map within the optical pupil diameter of the Si-side of the PMADM. A maximum displacement of $5.7 \times 10^{-6}$ m is observed when the separation between the magnets and the PMADM sample is 12 mm.

take the same deformation shape as that of Si-layer. A maximum displacement of around 5.76 $\mu$m is obtained at the Si-layer when the separation between the sample and the magnet is 12 mm. Figure 6(b) shows the simulated displacement map at the Si-layer side of the PMADM within an optical pupil diameter of 8 mm. The surface map is obtained by postprocessing the COMSOL displacement map in SAGUARO.

Figure 7(a) shows the simulation of maximum displacement of the PMADM surface with the varying separation distance between the magnets and the PMADM. A uniform magnetic field has been assumed across the face of the mirror and its strength falls down as the separation between the magnet and the mirror increases. As the separation distance between the magnets and the PMADM sample increases, the magnitude of deformation decreases due to a decrease in the magnetic field strength leading to less magnetic loading. Figure 7(b) shows the Zernike power coefficient obtained from postprocessing the simulated displacement maps in SAGUARO.

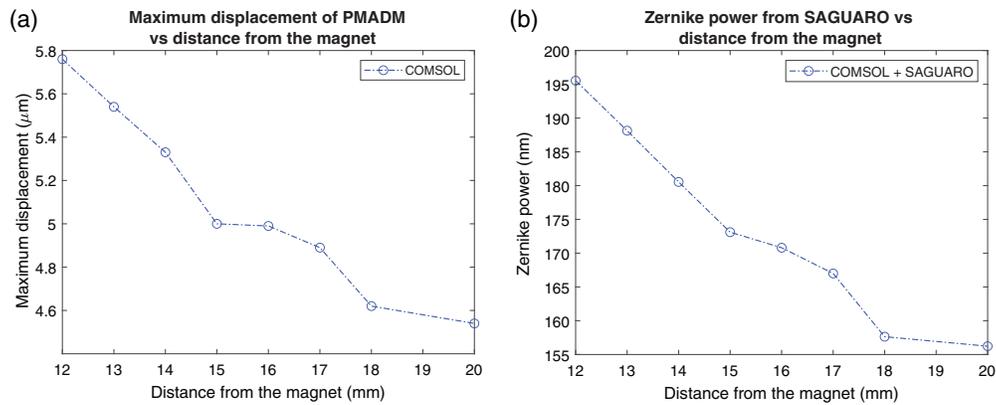

**Fig. 7** (a) Plot showing simulation of maximum displacement with changing separation distance between the magnet and the PMADM. (b) Zernike power term plot obtained from postprocessing the displacement maps obtained from COMSOL in SAGUARO. An optical pupil diameter of 8 mm is considered in the calculation.

The Zernike power term is calculated using SAGUARO by considering an optical pupil diameter of 8 mm. We observed that the power term decreases as the separation between the magnets and the PMADM sample increases. A decrease in magnetic loading causes less peak deformation, contributing to a smaller Zernike power coefficient.

### 3.2 Simulation of Laser Heating and Relaxation of PMADM

The temperature profile of the PMADM surface has been shown in Fig. 8(a) when heated by a laser with incident power of 50 mW. We can observe that the temperature within the optical pupil of the PMADM increases up to 307 K causing a change in magnetization that results in the relaxation seen in Fig. 8(b). With an increase from ambient temperature (303 K) to 307 K, the maximum displacement relaxes from 5.7 to 3.7 $\mu$m.

Figure 9(a) shows that with increasing laser power, the temperature of our PMADM substrate increases almost linearly. Figure 9(b) shows that with increasing temperature, the PMADM shows more relaxation in deflection, as an increase in temperature with increasing laser power causes a decrease in magnetization and magnetic load.

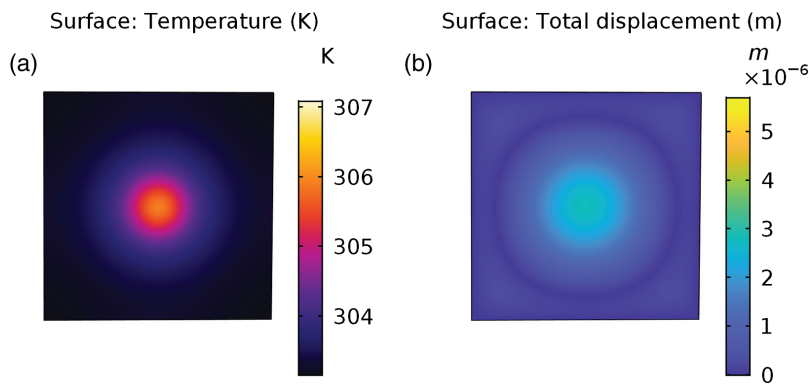

**Fig. 8** (a) Simulated temperature map of the P-layer of the PMADM when the distance between the magnets and the PMADM is 12 mm and heated by incident laser power of 50 mW at the magnetic PDMS side (P-layer). A maximum temperature of 307 K is generated as a result of the laser heating on the PMADM surface. (b) The resultant displacement map at the Si-layer of the PMADM due to relaxation by laser heating in presence of magnetic loading is shown. The maximum displacement relaxes to about $3.7 \times 10^{-6}$ m from $5.7 \times 10^{-6}$ m.

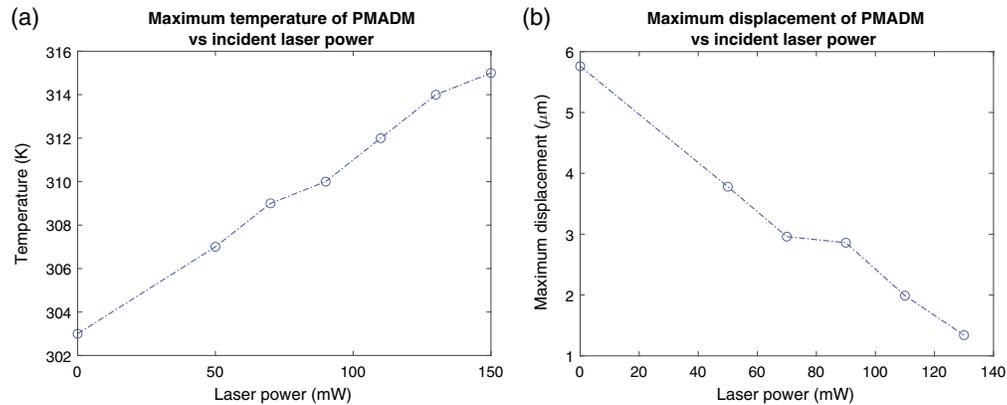

**Fig. 9** (a) Simulated PMADM temperature (measured at the P-layer) with varying laser power. (b) Maximum displacement due to both magnetic loading and laser heating at the Si-layer of the PMADM at different incident laser powers. As the laser power increases, more heating causes a temperature increase, which in turn, causes a decrease in magnetization and magnetic loading that results in more relaxation.

### 3.3 Experimental Test of Magnetic Deformation of PMADM

Figure 10 shows the displacement map (a), Zernike power term (b), and the residual surface map after 11 Zernike subtractions (c). These plots are obtained from experimental testing of the PMADM using a 4D interferometer when the PMADM is 12 mm from the magnets. A Zernike power coefficient of 195.5 nm is measured for this case. Bias terms, piston, and tip/tilt have been ignored. In Fig. 11, we have shown the change in Zernike coefficients of the power (focus), coma, astigmatism, and spherical aberration. Only the Zernike power coefficient term shows a decreasing trend, changing by more than 40 nm, whereas the other terms are stable to a maximum of 2.4 nm root-mean-squared error (RMSE). This is expected, given that the contribution to other Zernike terms is possibly due to initial stress in the optics induced during the mounting or fabrication stages. The power term is decreasing due to lesser peak deformation of the PMADM as the magnetic loading decreases with increasing separation between the magnet and the PMADM surface.

Figure 11 also shows the surface RMS values obtained from the 4D interferometer measurements after removing the 11 Zernike terms from the displacement map in SAGUARO. A maximum value of about 21 nm surface RMS is obtained, which is a measure of the quality of the optical substrate. Since the surface RMS is well within the $\lambda/8$ range ($\lambda = 632.8$ nm), the optical substrate layer can function as an optical mirror.

### 3.4 Comparison of Simulation and Experimental Results

In Fig. 12(a), we have compared the plots of the Zernike power coefficients obtained from the simulated and experimentally obtained displacement maps of the PMADM. The experimentally obtained displacement maps have been postprocessed in 4SIGHT and then in SAGUARO. The above step allows us to check the efficiency of SAGUARO in evaluating Zernike terms. This step is necessary as we cannot process our simulation data in 4SIGHT (a 4D software that analyzes interferometer data). The Zernike power coefficients obtained from processing experimental results in SAGUARO have an RMSE of 1.99 nm (1.11%) compared to that obtained from 4SIGHT. Since the RMSE is small, we can use SAGUARO to calculate Zernike terms from COMSOL simulation data and compare them with those obtained from experimental data using 4SIGHT. The above step will then allow us to check the validity of our FEM simulation model.

We have found the RMSE between the Zernike power terms obtained from postprocessing simulated displacement maps in SAGUARO and experimental displacement maps in 4SIGHT to be 1.54 nm (0.096%). We have fitted a curve on the Zernike power terms obtained from processing simulation data in SAGUARO that predicts their trend with varying separation between the magnet and the PMADM. The fitted curve has a 1.84-nm RMSE (1.12%) against

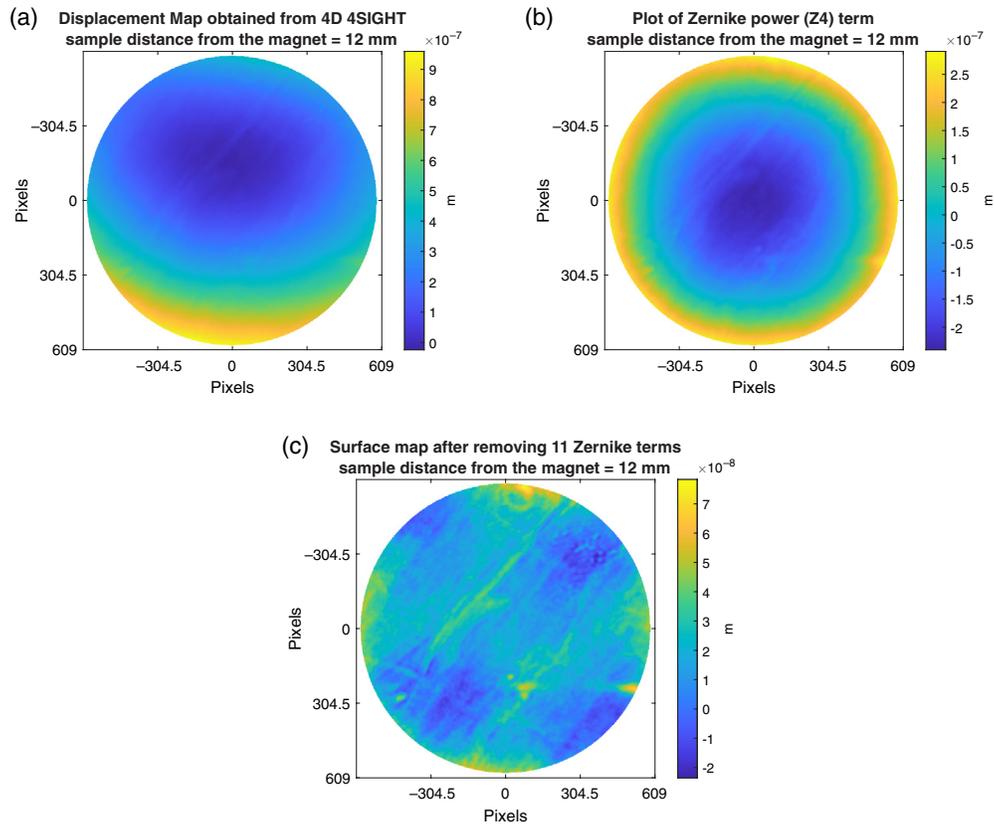

**Fig. 10** (a) Displacement map obtained from experimental testing of the Si-layer of the PMADM sample using a 4D interferometer[51] when the sample is 12 mm from the magnet. (b) Zernike power terms obtained after postprocessing the displacement map in SAGUARO. (c) To quantify the residual roughness, a map after subtracting 11 Zernike terms is shown with surface RMS of ∼21 nm. Residual mounting stress is seen through a four-sided pattern.

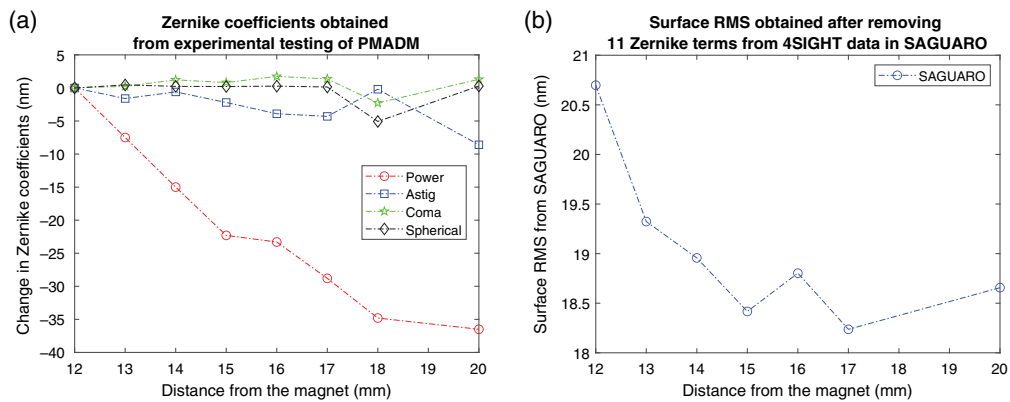

**Fig. 11** (a) Trend of Zernike coefficient terms obtained from experimental testing of the PMADM in the laboratory. A maximum change of 40 nm can be seen in the Zernike power terms when the separation between the magnet and the PMADM is increased from 12 to 18 mm. Other Zernike terms are stable to a maximum RMSE of 2.4 nm. (b) Surface RMS figure of the Si-layer side of the PMADM is also shown. A maximum value of 21 nm is calculated, which makes our optical substrate functional as a mirror surface.

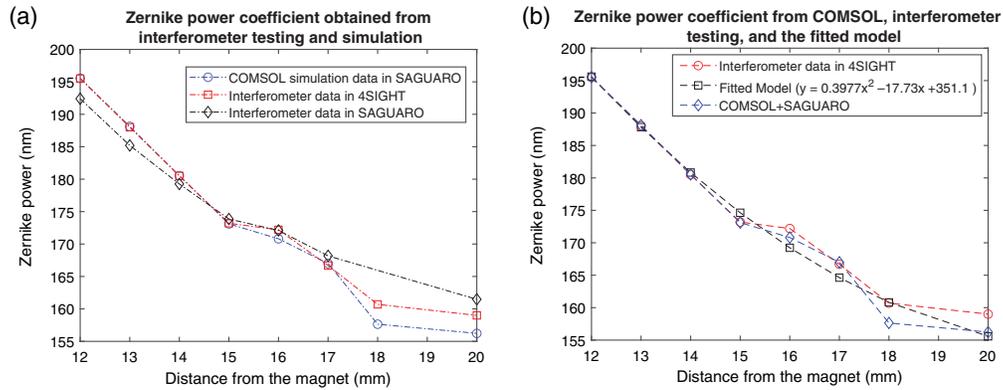

**Fig. 12** (a) Comparison between Zernike power term coefficients obtained from experimental laboratory testing (4SIGHT and SAGUARO) and our FEM model (COMSOL + SAGUARO) for our PMADM. (b) A curve fitted with 1.8382 nm RMSE (1.12%) predicts the nonlinear trend of the Zernike power term coefficient with varying separation distance between the magnets and the PMADM when the laser is off.

the experimental 4SIGHT data. The curve shows a nonlinear decreasing trend [shown in Fig. 12(b)] of the Zernike power term with the increasing separation distance between the magnets and the PMADM. This behavior is expected due to the decreasing magnetic field strength with increasing separation between the magnets and the PMADM. The trend is governed by a second-order polynomial (shown in the same figure) due to the inverse relationship between the magnetic field strength and the separation distance. The fitted polynomial provides a good approximation of the PMADM model to calculate the Zernike power term. However, our FEM model should obtain a more accurate estimation as it shows the least RMSE compared to the experimental 4SIGHT data.

### 3.5 *Simulation of Maximum Stroke*

The stroke or the maximum surface change, which can be applied to a device before failure is a key parameter for design of active optical systems. Figure 13(a) shows the von Mises stress generated in the PMADM when the separation between the magnets and the PMADM is 12 mm and the laser is off. We have also calculated the maximum stress generated in the PMADM using our COMSOL simulation model for varying separation distances between the magnets and the PMADM. Maximum stroke calculations seen in Fig. 13(b) takes into consideration a 3× safety margin, achieved by comparing the yield strength of the PDMS (which is lowest among the

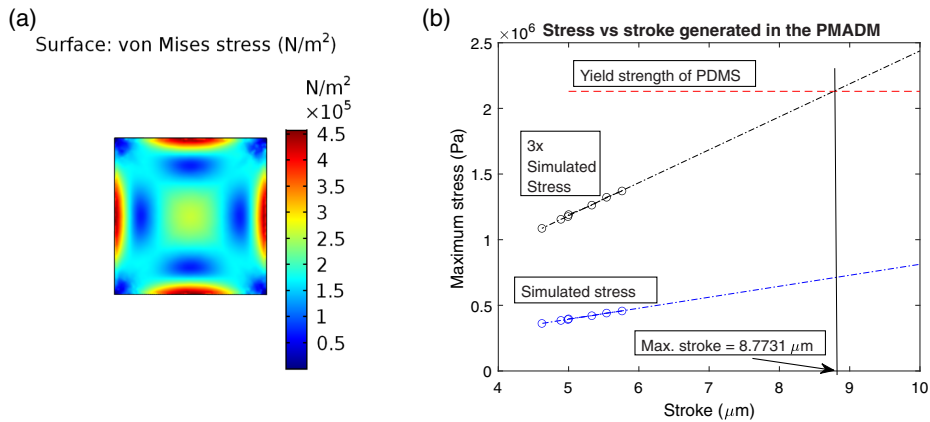

**Fig. 13** (a) von Mises stress generated in the PMADM when the separation between the magnets and the PMADM is 12 mm. (b) Simulated stress and calculation of maximum stroke with 3× safety margin.

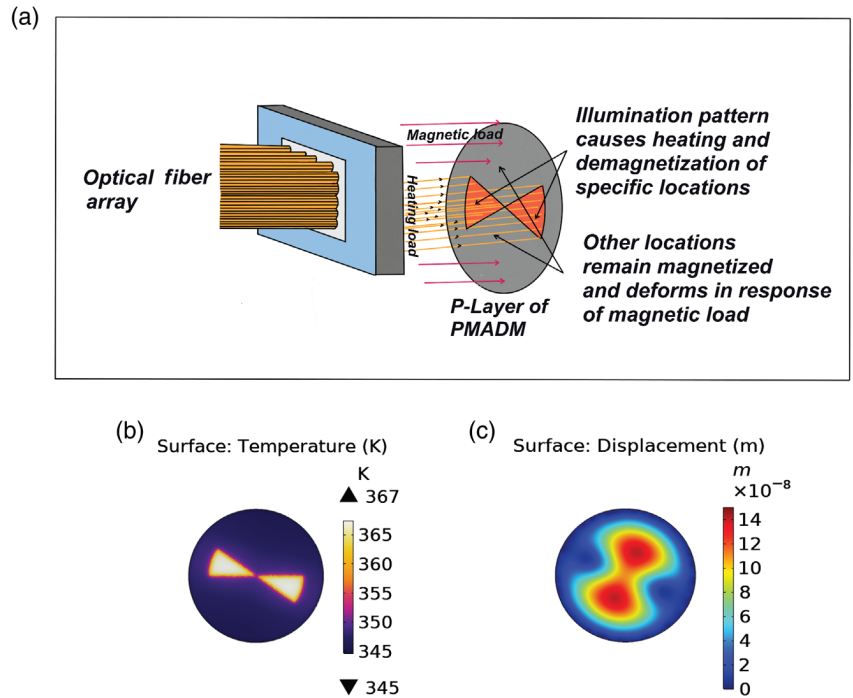

**Fig. 14** (a) Illustration showing how an illumination pattern can be generated using grids of optical fiber for heating and demagnetization of PMADM. (b) Simulated temperature profile generated at the P-layer of the PMADM when illuminated by 120 mW power over 4.8 mm$^2$ illuminated area. (c) Astigmatism like surface deformation generated at the Si-layer of the PMADM because of the illumination and magnetic loading. The distance between the magnet and the PMADM sample is 17 mm for this simulation with magnet's south pole facing the P-layer of the PMADM.

materials used in fabricating the PMADM) with three times the maximum stress generated in the PMADM at a different stroke. The maximum stroke is identified as the point where the 3× stress matches the yield strength of the PDMS. From our simulation results, we have estimated a maximum stroke of around 8.78 $\mu$m for our PMADM model before risk/failure with a 3× safety margin.

### 3.6 Simulation of Higher Order Control

We have also performed a COMSOL simulation study where the P-layer of the PMADM is illuminated in a specific pattern to obtain astigmatism like deformation at the Si-layer of the PMADM when the P-layer of the sample is kept at a distance of 17 mm facing the south pole of the magnets. A total power of 120 mW is deposited over the illuminated pattern covering 4.8 mm$^2$ area. Such an illumination pattern can be easily achieved in laboratory setup via a compact optical fiber grid.[54] This simulation study also hints at the capability of the PMADM in aberration corrections beyond focus control. Figure 14(b) shows the illumination pattern and the temperature distribution due to 120 mW heating at the P-layer of the PMADM. A maximum temperature of 367 K is reached at the P-layer of the PMADM. Figure 14(c) shows the surface deformation profile at the Si-layer of the PMADM created due to the magnetic loading and illumination. A maximum deformation of about 150 nm is achieved. From this study, it is clear that the area that is heated by the illumination pattern gets demagnetized and hence deforms less compared to the other areas within the optical pupil of the Si-layer of the PMADM.

## 4 Summary and Future Work

This work presents laboratory testing of a prototype PMADM and its COMSOL simulation model, which demonstrates focus adjustment using a magnet and a laser heating source.

A spatially varying magnetic field or array of magnets would provide an alternative actuation mechanism for the presented device, though the precision afforded by zonal photodemagnetization is expected to provide superior spatial resolution and actuation precision. The presented prototype of our PMADM allows control of focus and can also be modified to correct other aberrations. To correct higher-order aberrations, an array of optical fibers with specific illumination pattern may be used to provide addressable optical control using our PMADM. In the future, we will also explore higher surface quality substrates[55] (>1 RMS), including higher quality silicon surfaces. The PMADM has potential applications in the field of precision wavefront correction and allows autofocus adjustment for imaging and optomechanics purposes. It can also be used for correcting slow low-order wavefront errors[56,57] (<1 Hz) that may originate due to attitude control systems and thermal changes in the optics of a space-based observatory. Ground-based applications of active or adaptive optics on slower time scales include correcting gravitational sag, differential flexure, and thermal drift.[58,59] We are especially looking forward to studying the PMADM's response to the laboratory laser heating test, which will provide demagnetization and focus adjustment without translation of the drive magnet. Since photoactuation requires heating to achieve demagnetization, thermal relaxations times for these actuators are expected to be relatively slow (∼1 Hz) (Li et al.). However, improvements can be made by reducing the PDMS-C layer thickness or by increasing the heat transfer coefficient as shown in Fig. 18 (see Appendix C). Using a thin, homogeneous, metallic, magnetic substrate can also aid in conduction in vacuum environments, thus enhancing the response time of the PMADM.

Our PMADM optics are not initially flat and while the current implementation presents with an initial curvature, this is not unusual in active optical systems, where the residual stress curvature of MEMS DMs[60,61] is commonly accounted for by an initial defocus of upstream optics or a "flat-map" of distortion voltages.

Additionally, focus powered DMs are desired to minimize the number of reflections in some systems, for example, in coronagraphs[62] and adaptive optics systems.[31] Mounting geometry optimization and bonding process refinement are also expected to minimize the mounting stress visible in Fig. 10.

The main advantage of our PMADM is that it performs focus adjustment in a complete contactless fashion and can be tuned to have a large or small stroke to correct other aberrations. Future work also includes using an electromagnet to test the device till failure to validate the maximum stroke simulation presented in Fig. 13.

The prototype can also be scaled to larger areas using spring actuators as shown in Fig. 15. A larger area will require less magnetic load for the same amount of focus change and the spring actuators will help in keeping the mirrors flat under magnetic load. Local demagnetization will allow springs to relax locally in comparison to other spring actuators and will offer more spatial control.

The technology that we have presented shows potential for applications where precision is required and where optical control is more feasible than using high voltages/powers, such as in vacuum, space, and low-gravity environments.

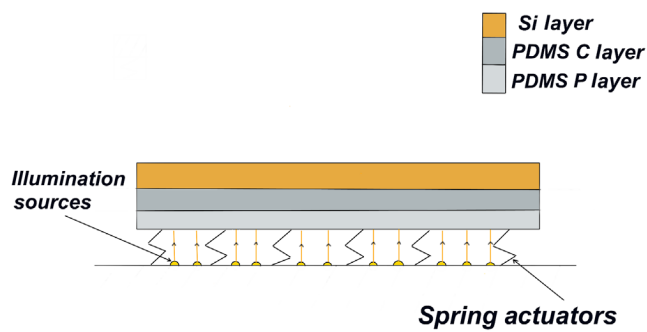

**Fig. 15** Illustration showing how the PMADM can be supported using spring actuators to facilitate scaling to larger areas. The magnetic load will compress the springs and will help in keeping the mirrors flat. Local demagnetization will allow relaxation of springs locally providing more spatial control. Demagnetization of the magnetic layer can be achieved using illumination sources as shown in the illustration.

## 5 Appendix A: Simulation of KJ DC2E Magnets

Figure 16(a) shows the magnetic flux density plot of a stack of eight KJ DC2E disk magnets simulated in COMSOL. Figure 16(b) shows the comparison plot between the magnetic field strength obtained from KJ Magnetics and our COMSOL simulation model with varying probe distance from the magnet. The RMSE obtained is 104.77 G (17%), which shows that our FEM model successfully simulates the magnetic field of the eight KJ DC2E disk magnet stack with varying probe distance from the magnet. The magnetic properties and parameters used for modeling the magnets in COMSOL are given in Table 2.

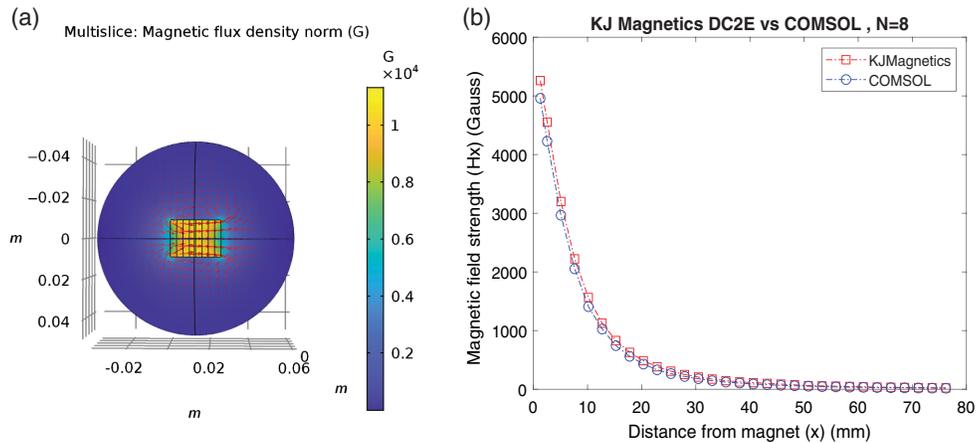

**Fig. 16** (a) Plot of magnetic flux density of eight KJ DC2E magnets in COMSOL. (b) Comparison plot of varying magnetic field strength with varying probe distance from the magnet. The RMSE percent error is 17%.

**Table 2** Specifications and magnetic properties of the eight KJ DC2E magnets used in our model.

| Specification | Neodymium magnet |
| --- | --- |
| Model | DC2E KJ magnetics |
| Geometry type | Cylindrical (disc) |
| Radius (mm) | 9.53 |
| Thickness (mm) | 3.18 |
| Remnant flux density (Gauss) | 13,200 |
| Relative permeability | 1.05 |

## 6 Appendix B: Simulation of Magnetic Properties of Magnetic PDMS Layer

Figure 17(a) shows the hysteretic magnetization plot of pretreated $CrO_2$ by Li et al., whereas Fig. 17(b) shows the relative permeability ($\mu$) values that we have used in our FEM model to compensate for the nonlinear variation of the magnetization of the magnetic PDMS layer with changing applied magnetic field. This compensation has been made because, as is clear from Fig. 17(a), the slope of the magnetization curve with changing magnetic field strength is not constant. As the magnetic field strength decreases, the slope of the hysteretic magnetization curve increases, which is directly proportional to the magnetic susceptibility ($\chi$). The magnetic susceptibility is related to relative permeability and to accommodate these changes, we have used relative permeability as a variable in our FEM model. This accounts for the adjustment in the $\chi$ values and improves the accuracy of the FEM model.

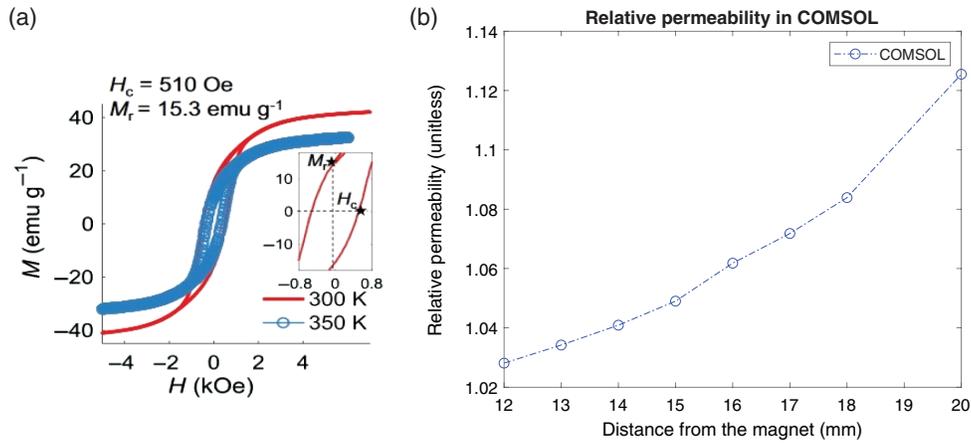

**Fig. 17** (a) Hysteretic plot of magnetization of $CrO_2$ by Li et al. (b) Plot of relative permeability with changing separation distance from the magnet.

## 7 Appendix C: Simulation of Relaxation Time of PMADM

The relaxation time of the PMADM can be defined as the time taken by the magnetic side of the prototype to regain its magnetization via fading of the demagnetization effect. Since the magnetization of the magnetic layer of the PMADM (PDMS C layer) is proportional to its temperature, the relaxation time is equivalent to the thermal relaxation time of the PDMS C layer. Hence, the relaxation time can be defined by Eq. (5) where $\tau$ is the relaxation time (s), $\rho$ is the density of the PDMS C layer, $C$ is the heat capapcity of the PDMS C layer, $b$ is the thickness of the PDMS C layer, and $h$ is the convection coefficient:

$$\tau = \rho C b / h. \tag{5}$$

Using Eq. (5), we can see that the relaxation time (in seconds) is directly proportional to the thickness of the PDMS C layer, whereas it is inversely proportional to the convection coefficient.

Improvements in relaxation time can be made by reducing the thickness of the PDMS C layer or by increasing the convection coefficient. Figure 18 shows the simulated plot of the relaxation time of the PMADM prototype versus the thickness of its PDMS C layer (a) and with the convection coefficient (b). We can deduce from the plot that a significant improvement in the relaxation time can be made by increasing the convection coefficient. The relaxation time can also be improved by introducing a thin metallic magnetic substrate, which will aid in conduction in vacuum environments, thus improving the response time. Introducing cold bias, a relatively lower surrounding temperature can also help in escalating the process of heat transfer from the PMADM to the environment and will help in further improving the response time.

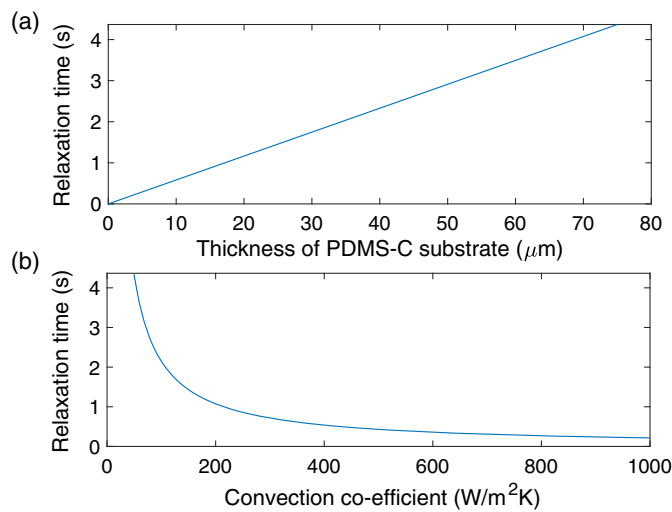

**Fig. 18** Simulation of relaxation time of PMADM for (a) different thicknesses of PDMS-C layer and (b) heat convection coefficient. The relaxation time can be improved linearly by reducing the thickness of the PDMS-C layer or can be improved exponentially by increasing the heat convection coefficient.

## Acknowledgments

The authors would like to acknowledge the feedback of the University of Arizona Space Astrophysics Lab (UASAL) members and all the friends and family (including pets Ash and Lina) who uplifted our spirits during the COVID-19 pandemic. Writing assistance was provided by Nicole Miklus. Portions of this work were supported by the Arizona Board of Regents Technology Research Initiative Fund (TRIF). This research made use of the high-performance computing resources supported by the University of Arizona (UA) TRIF, UITS, and RDI and maintained by the UA Research Technologies Department. The authors declare no conflicts of interest.

## Code, Data, and Materials Availability

Simulation model and experimental data underlying the results will be provided upon reasonable request.

**Amit Kumar Jha** received his BTech degree in electronics and instrumentation engineering from Techno Main affiliated to West Bengal University of Technology in 2017. He is a PhD student at Wyant College of Optical Sciences of the University of Arizona. His research interests are in the area of astronomical instrumentation and quantum imaging and sensing.

**Ewan S. Douglas** received his bachelor's degree in physics from Tufts University in 2008 and his master's and doctorate degrees in astronomy from Boston University in 2011 and 2016, respectively. He completed his postdoctoral appointment at Massachusetts Institute of Technology, where he served as a DeMi payload engineer. He is an assistant professor of astronomy at the University of Arizona (Steward Observatory). His research focuses on space instrumentation, wavefront sensing and control, and high-contrast imaging of extrasolar planets and debris disks.

**Meng Li** received her bachelor's degree of engineering from the Southeast University, Nanjing, China, her master's degree of science from the Advanced Material Program in the University of Ulm, Ulm, Germany, and her PhD in biomedical engineering in 2020 from Tufts University, Medford, Massachusetts, USA. She is a postdoctoral researcher at Max-Planck-Institute for Intelligent Systems, Stuttgart, Germany, and a fellowship receiver at Alexander von Humboldt

Foundation. Her research interests are in developing new functional materials and systems for soft robots and actuators.

**Corey Fucetola** received his SB degree in mathematics and SB, MEng, and PhD degrees in electrical engineering and computer science from MIT, where he also did his postgraduate work. He is the cofounder of Sublamit LLC. He is a research assistant professor in biomedical engineering at Tufts University.

**Fiorenzo G. Omenetto** is the Frank C. Doble professor of engineering and a professor of biomedical engineering at Tufts University. He is the dean of the research for the School of Engineering and holds secondary appointments in the Department of Physics and the Department of Electrical Engineering. His research interests are the interface of technology, biologically inspired materials, and the natural sciences, with an emphasis on new transformative approaches for sustainable materials for high-technology applications.